\documentclass[twocolumn,showpacs,amssymb,aps,prd,floatfix]{revtex4-1}

\usepackage{epsfig}

\newcommand{\be}{\begin{equation}}
\newcommand{\ee}{\end{equation}}
\newcommand{\bea}{\begin{eqnarray}}
\newcommand{\eea}{\end{eqnarray}}

\newcommand{\xv}{{\mathbf x}}
\newcommand{\vecnul}{{\mathbf 0}}
\newcommand{\bra}{\langle}
\newcommand{\ket}{\rangle}

\newcommand{\eps}{\epsilon}
\newcommand{\om}{\omega}

\newcommand{\cC}{${\cal C}\;$}

\newcommand{\half}{\frac{1}{2}}

\begin{document}

\title{Nucleons and parity doubling across the deconfinement transition}

\author{Gert Aarts$^a$}
\author{Chris Allton$^a$}
\author{Simon Hands$^a$}
\author{Benjamin J\"ager$^a$}
\author{Chrisanthi Praki$^a$}
\author{Jon-Ivar Skullerud$^b$}

\affiliation{$^a$Department of Physics, College of Science, Swansea University, Swansea SA2 8PP, United Kingdom}
\affiliation{$^b$Department of Mathematical Physics, National University of Ireland Maynooth, Maynooth, County Kildare, Ireland}

\date{June 25, 2015}

\begin{abstract} 
It is expected that nucleons and their parity partners become degenerate when chiral symmetry is restored. We investigate this question in the context of the thermal transition from the hadronic phase to the quark-gluon plasma, using lattice QCD simulations with $N_f=2+1$ flavours. We observe a clear sign of parity doubling in the quark-gluon plasma. Besides, we find that the nucleon ground state is, within the uncertainty, largely independent of the temperature, whereas temperature effects are substantial in the negative-parity ($N^*$) channel, already in the confined phase. 
  
\end{abstract}


\maketitle


\section{Introduction}

The role of discrete and continuous symmetries played a fundamental role in the development of the theory of the strong interaction, Quantum Chromodynamics. Chiral symmetry breaking and its restoration remain topical subjects, mostly due to the creation of the quark-gluon plasma at relativistic heavy-ion collision experiments at the Large Hadron Collider (CERN) and the Relativistic Heavy Ion Collider (BNL). It is expected that chiral symmetry will be restored at high temperature, as seen e.g.\ in nonperturbative studies using lattice QCD simulations \cite{Borsanyi:2010bp,Bazavov:2011nk}. 

In the past decades chiral symmetry restoration at finite temperature has been studied in great detail in the mesonic sector \cite{Rapp:1999ej}. One reason is that mesonic correlation functions are relatively easily accessible on the lattice \cite{Karsch:2003jg,Ding:2012ar}. Moreover, susceptibilities related by chiral symmetry, such as in the pion and scalar meson channels, can now be computed using chiral lattice fermions \cite{Bhattacharya:2014ara}.
Unlike the mesonic sector, the baryonic sector has hardly been investigated at finite temperature (early work on screening masses from lattice QCD can be found in Ref.\ \cite{DeTar:1987ar}, and, in the presence of a small chemical potential, in Ref.\ \cite{Pushkina:2004wa}). 
Nevertheless, understanding the behaviour of nucleons in a hadronic medium or in the quark-gluon plasma is relevant for heavy-ion collisions, where proton spectra  are routinely measured and compared to theoretical predictions. Just as for mesons, possible in-medium modification of nucleons and other baryons might affect  signals observed in those experiments.

In the baryon sector the combination of chiral symmetry and parity leads to a prediction readily testable in QCD:  namely
that of parity doubling, i.e., a degeneracy between channels related by parity, provided that both symmetries are realised (the argument will be briefly reviewed below). At zero temperature, where chiral symmetry is broken, parity doubling is not observed, except perhaps in the case of excited hadrons \cite{Glozman:2007ek}. 
However, since chiral symmetry is restored at high temperature, it should become relevant in the quark-gluon plasma. 

Recently the question of parity doubling has been taken up in Ref.\ \cite{Datta:2012fz}, where it was studied at three temperatures in quenched lattice QCD. 
In this paper we present what is, as far as we know,  the first study of nucleons at finite temperature in lattice QCD with $N_f=2+1$ dynamical quarks, for a range of temperatures below and above the deconfinement transition. We find clear indications of parity doubling, occurring in coincidence with the deconfinement crossover. Moreover, within our numerical uncertainty, the mass of the nucleon ground state is found to be independent of the temperature of the hadronic medium.

\section{Nucleon propagation} 

The standard interpolation operator for a nucleon, which we will consider below, is given by (the material reviewed here is well-known, see e.g.\ the textbooks \cite{Montvay:1994cy,Gattringer:2010zz})
\be
O_N(\xv,\tau) = \eps_{abc} u_a(\xv,\tau) \left[ u_b^T(\xv,\tau) {\cal C} \gamma_5 d_c(\xv,\tau)  \right],
\ee
where $u, d$ are the quark fields,  $a,b,c$ are colour indices, other indices are suppressed and \cC denotes the charge conjugation matrix. Under parity one finds that 
\be
{\cal P} O_N(\xv,\tau) {\cal P} =\gamma_4 O_N(-\xv,\tau),
\ee
and hence operators for the positive and negative parity channels are obtained as
\be
\label{eq:Opm}
O_{N_\pm}(\xv,\tau) = P_\pm O_N(\xv,\tau),
\quad\quad
P_\pm=\half(1\pm\gamma_4).
\ee
We consider the usual euclidean correlators, summed over the Dirac indices and projected to zero momentum,
\be
G_\pm(\tau) = \int d^3x\, \left\bra O_{N_\pm}(\xv,\tau) \overline{O}_{N_\pm}(\vecnul,0) \right\ket.
\ee
It follows from the properties under euclidean time reflection that, in the case of $G_+(\tau)$, forward (backward) propagation in time corresponds to the positive-parity (negative-parity) channel. On a lattice at a nonzero temperature $T$, with $0\leq\tau<1/T$, the negative-parity channel is  then propagating with $\tau_-=1/T-\tau$. 
Hence both parity channels can be obtained from the same correlator, either $G_+(\tau)$ or $G_-(\tau)$.
In the case that the signal is dominated by the ground states in both channels, this leads to the simple exponential Ansatz, 
\be
\label{eq:Ansatz}
G_\pm(\tau) = A_\pm e^{-m_\pm\tau} +  A_\mp e^{-m_\mp(1/T-\tau)},
\ee
with two masses $m_\pm$. 
We note that the spectral representation is  slightly more complicated than for mesonic correlators, and reads
\be
 G_\pm(\tau)  = \! \int_{-\infty}^\infty \frac{d\om}{2\pi} \! \left[  \frac{e^{-\om\tau}}{1+e^{-\om/T}} \rho_\pm(\om) -  \frac{e^{\om(\tau-1/T)}}{1+e^{-\om/T}} \rho_\mp(\om)  \right],
\label{eq:spec}
\ee
where $\rho_\pm(\om)$ are the spectral functions in the positive and negative parity channels \cite{prep}.

Provided that chiral symmetry is unbroken, performing a chiral rotation on the quark fields immediately leads to the result that the two parity channels are degenerate, and \cite{Gattringer:2010zz}
\be
 G_\pm(\tau) = G_\mp(\tau) = G_\pm(1/T-\tau),
\ee
up to overall minus signs. 
In nature, chiral symmetry is broken at zero temperature. Indeed, the ground states in the two parity channels differ substantially in mass, with the nucleon $N(939)$ considerably lighter than the negative-parity partner $N^*(1535)$. Here we study what happens when the temperature is increased and chiral symmetry is eventually no longer spontaneously broken. We only consider $G_+(\tau)$ and drop the $+$ from now on.

\begin{table}[t]
\begin{center}
\begin{tabular}{ccccccc}
    \hline\hline
      $N_s$ & $N_\tau$  & $T$  [MeV] & $T/T_c$  & $N_{\rm cfg}$  & $N_{\rm src}$ \\
\hline\hline
  24&128 &  44	 & 0.24 & 171   & 2\\
  24 & 40 & 141 & 0.76 &  301  & 4 \\
  24 & 36 & 156 & 0.84 &  252  & 4 \\
  24 & 32 & 176 & 0.95 & 1000 & 2 \\ 
  24 & 28 & 201 & 1.09 &   501 & 4 \\ 
  24 & 24 & 235 & 1.27 & 1001 & 2 \\  
  24 & 20 & 281 & 1.52 & 1000 & 2 \\ 
  24 & 16 & 352 & 1.90 & 1001 & 2 \\ 
      \hline\hline
   \end{tabular}
\caption{Details of the ensembles. The lattice size is $N_s^3\times N_\tau$, with the temperature $T=1/(a_\tau N_\tau)$. 
  $N_{\rm cfg}$ ($N_{\rm src}$) denotes the number of configurations (sources) used at each volume. 
    The spatial lattice spacing is $a_s=0.1227(8)$ fm, with anisotropy $a_s/a_\tau=3.5$.
  }
\label{tab:lat}
\end{center}
\end{table}

\section{Lattice details} 

 We study this question using lattice QCD simulations with $N_f=2+1$ flavours on anisotropic lattices, with a smaller temporal lattice spacing $a_\tau<a_s$, namely $a_s/a_\tau=3.5$.  We use a Symanzik-improved anisotropic gauge action with tree-level tadpole coefficients and a tadpole-improved Wilson-clover fermion action with stout-smeared links;  details on the lattice action and parameters at zero temperature can be found in the work of the Hadron Spectrum Collaboration \cite{Edwards:2008ja}. The strange quark mass is tuned to the physical value, while the light quarks correspond to $M_\pi=384(4)$ MeV, with $M_\pi/M_\rho=0.466(3)$ \cite{Lin:2008pr}.
  We consider a number of ensembles at nonzero temperature, as summarised in Table \ref{tab:lat}. The pseudo-critical temperature $T_c$ is determined via the renormalised Polyakov loop. 
 The larger value of $T_c$ than expected is primarily due to the light quarks being heavier than in nature.
 More details on the finite-temperature ensembles are available in Refs.\ \cite{Amato:2013naa,Aarts:2014nba}.

For the nucleon operators we employ the form (\ref{eq:Opm}), with Gaussian smearing for the sources $\eta$ and sinks, translated over the lattice, using \cite{Gusken:1989ad}
\be
\eta' = C\left(1+\kappa H\right)^n\eta, 
\ee
where $H$ is the spatial hopping part of the Dirac operator and $C$ an appropriate normalisation. Most of the results are obtained using $\kappa=8.7$ and $n=140$ \cite{Capitani:2012gj}, applying  the same smearing parameters at all temperatures. 
 The smearing procedure was tuned to maximise the length of the effective-mass plateau in the positive-parity nucleon channel at the lowest temperature.
 The links are APE smeared, using one iteration with $\alpha=1.33$ \cite{Albanese:1987ds}.
 Correlation functions are generated using the Chroma software suite \cite{Edwards:2004sx}.
 A discussion of smearing dependence is given at the end. More details will be available in Ref. \cite{prep}.

\begin{figure}[t]
\begin{center}
\includegraphics[width=0.98\columnwidth]{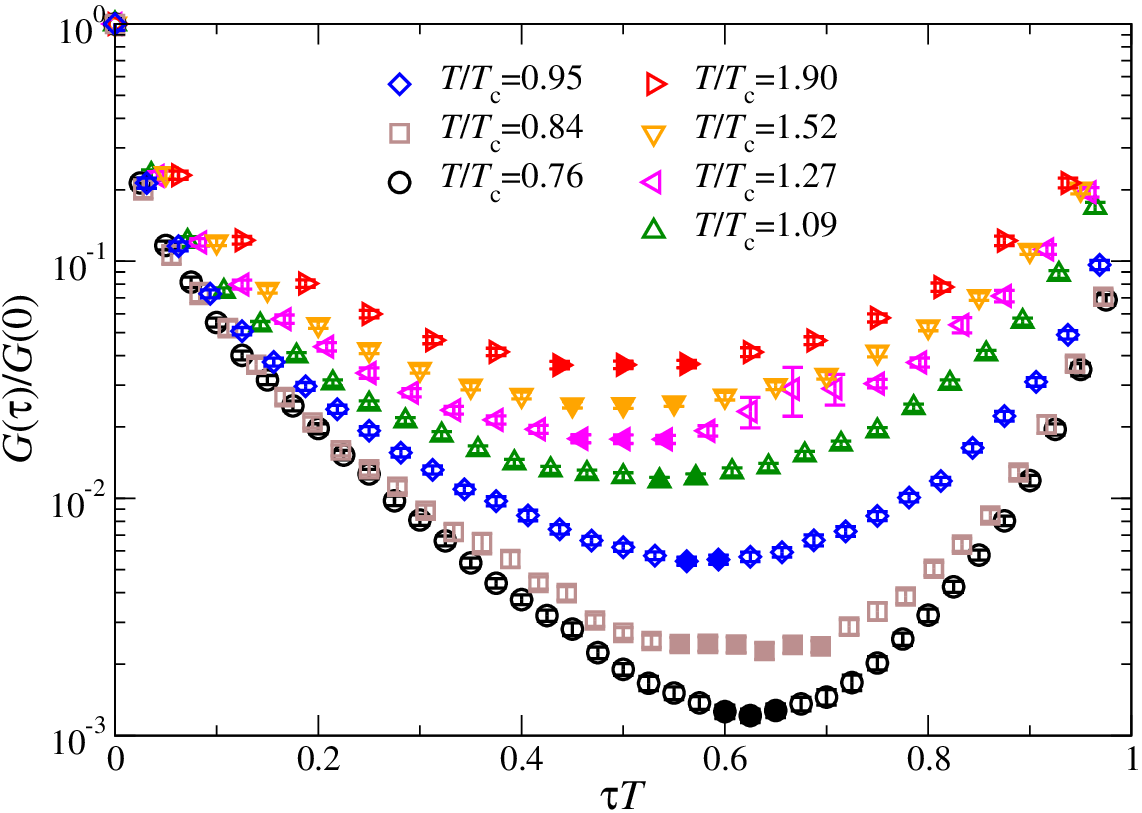}  
 \caption{Euclidean correlator $G(\tau)/G(0)$ as a function of $\tau T$. At each temperature, the filled symbols indicate the minimum of the correlator, within the error.
  }
\label{fig:Gtau}
\end{center}
\end{figure}

\section{Results}

The nucleon-nucleon correlators are shown in Fig.\ \ref{fig:Gtau} on a logarithmic scale as a function of $\tau T$, for the various temperatures we consider (the lowest temperature is not shown). In the hadronic phase we observe approximate exponential decay for both the forward and the backward propagating mode, but with a clear absence of reflection symmetry around $\tau T=1/2$, indicating that the positive-parity nucleon, with mass $m_+$, is considerably lighter than the negative-parity nucleon, with mass $m_-$. 
As the temperature is increased, the correlator becomes more and more symmetric; we have used filled symbols to indicate at which time slices the correlator is minimal, within the statistical error. This moves towards the centre of the lattice, indicating that forward and backward propagation become degenerate, i.e.\ we find parity doubling. 

\begin{figure}[t]
\begin{center}
\includegraphics[width=0.98\columnwidth]{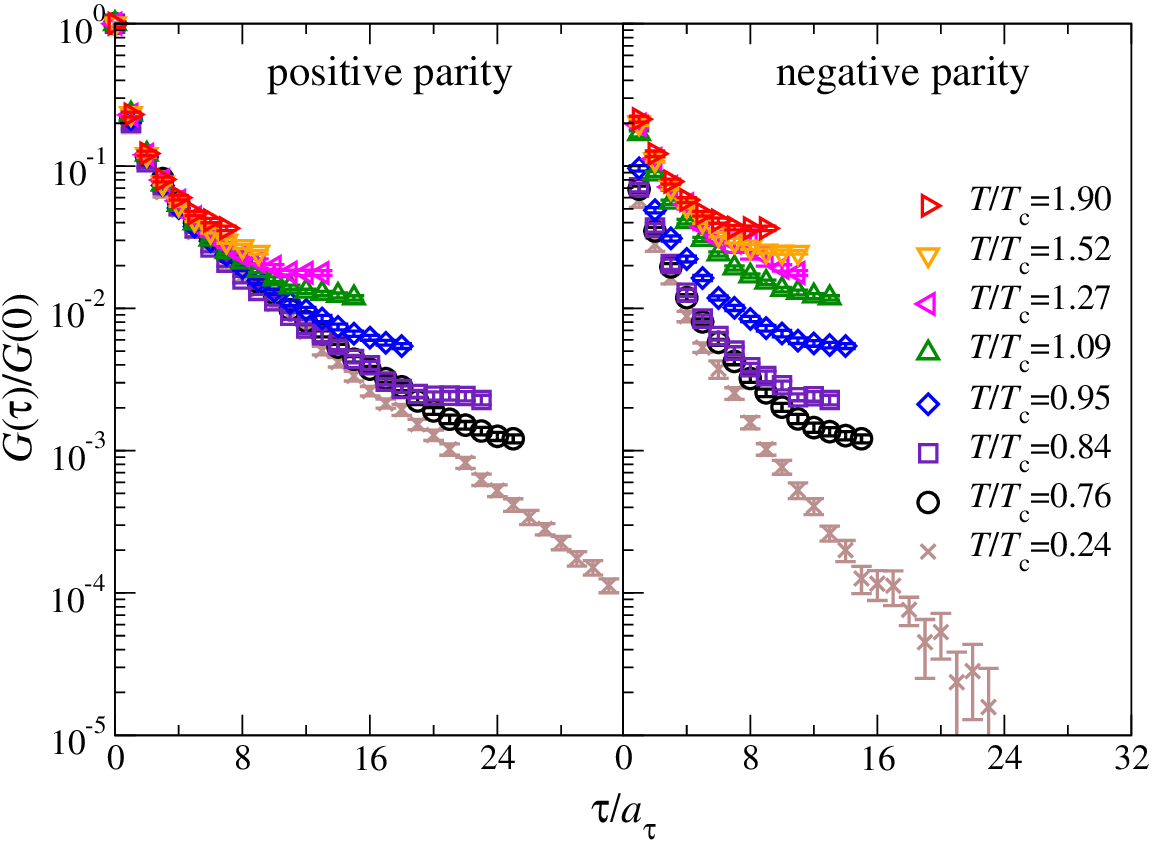}  
 \caption{Euclidean correlator $G(\tau)/G(0)$ versus $\tau/a_\tau$ for the positive-parity channel (left) and negative-parity channel, starting at the opposite side of the lattice (right). 
  }
\label{fig:Gtau-parity}
\end{center}
\end{figure}

To find the possible temperature dependence of the ground states, we show in Fig.\ \ref{fig:Gtau-parity}  both parity channels from the same correlator, but now as a function of $\tau/a_\tau$, starting at the opposite side of the lattice for the negative-parity channel. The lowest temperature is now also included.
Note that we show the correlator up to its minimum. On the positive-parity side, we observe considerably less temperature dependence than on the negative-parity side.
This is further demonstrated in Table II, where we show the results from fits to the exponential Ansatz (\ref{eq:Ansatz}) below $T_c$. 
The error combines estimates of statistical and systematic uncertainties in quadrature. At the lowest temperature, we note that both masses are higher than in nature (see also Refs.\ \cite{Lin:2008pr,Bulava:2010yg}), 
presumably due to the quarks being too heavy, but that the mass splitting $m_--m_+\sim 700$ MeV is of the right order, albeit with a large error. The positive-parity mass is, within the error, temperature independent, while $m_-$ shows significant temperature dependence.
We found that above $T_c$ simple exponential fits are no longer reliable. This may be due to the absence of a clear ground state, i.e.\ the nucleon is no longer a well-defined particle. 

 To study the onset  of parity doubling in the confined phase, the final column contains the dimensionless ratio
\be
\Delta=\frac{m_--m_+}{m_-+m_+},
\ee
 with estimates of statistical and systematic errors added in brackets. Note that in nature, $\Delta=0.241$ at zero temperature, while in the parity-degenerate case, $\Delta=0$. We observe a reduction of $\Delta$ as the temperature is increased, with a rather large systematic error arising mostly from the difficulty of determining the groundstate mass in the negative-parity channel at finite temperature.

\begin{table}[t]
\begin{center}
\begin{tabular}{cccccc}
    \hline\hline
      $T/T_c$ & $a_\tau m_+$ & $a_\tau m_-$  & $m_+$ [GeV]  & $m_-$ [GeV]  & $\Delta$\\
\hline\hline
   0.24 & 0.213(5)$\;\;$    & 0.33(5)   	& 1.20(3)$\;\;$     & 1.9(3) & 0.209(28)(082) \\
   0.76 & 0.209(16)  & 0.28(3)   		& 1.18(9)$\;\;$     & 1.6(2) & 0.138(29)(130) \\
   0.84 & 0.192(17)  & 0.28(2)   		& 1.08(9)$\;\;$     & 1.6(1) & 0.197(39)(054)  \\ 
   0.95 & 0.198(25)  & 0.22(4)   		& 1.12(14) 	  & 1.3(2)  & 0.052(35)(190)  \\      
   \hline\hline
   \end{tabular}
\caption{Results from exponential fits to Eq.~(\ref{eq:Ansatz}) below $T_c$, with statistical and systematic errors added in quadrature. The final column contains $\Delta=(m_--m_+)/(m_-+m_+)$ with estimates of statistical and systematic errors added in brackets. In nature, $\Delta=0.241$ at $T=0$.
  }
\label{tab:fit}
\end{center}
\end{table}

 In order to investigate parity doubling in more detail also above $T_c$, we consider the following ratio of correlation functions \cite{Datta:2012fz}
\be
\label{eq:Rt}
R(\tau) = \frac{ G(\tau)-G(1/T-\tau)}{ G(\tau) + G(1/T-\tau)}.
\ee
Note that $R(1/T-\tau)=-R(\tau)$ and hence $R(1/2T)=0$.
Consider first low temperature. In the (extreme) case that the correlator is dominated by the positive-parity ground state and that the negative-parity ground state is much heavier, $m_-\gg m_+$, we find that $R(\tau)=1$, except near $\tau T=1/2$, due to the reflection asymmetry. 
On the other hand, in the case of parity doubling, with $G(\tau)=G(1/T-\tau)$, we find that $R(\tau)=0$. Hence this ratio lies naturally between 0 and 1.

\begin{figure}[t]
\begin{center}
\includegraphics[width=0.98\columnwidth]{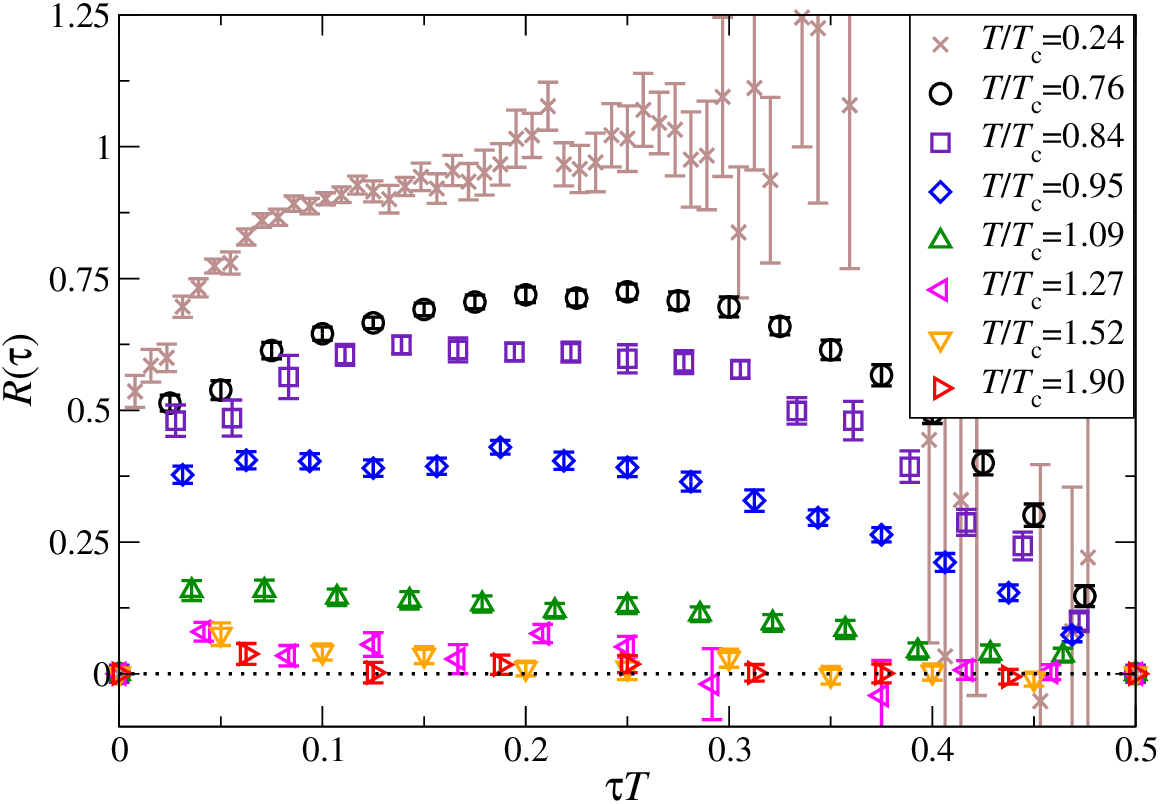}  
 \caption{Ratio $R(\tau)$ in Eq.\ (\ref{eq:Rt}) as a function of $\tau T$.
  }
\label{fig:Rtau}
\end{center}
\end{figure}

The results for $R(\tau)$ are shown in Fig.\ \ref{fig:Rtau}. We observe that the ratio is distinctly different from zero in the hadronic phase,
 indicating the absence of parity doubling, with an appreciably lighter ground state in the positive-parity channel. 
 The ratio is fairly constant across the entire euclidean time range: the drop  towards zero as $\tau T$ approaches $1/2$ follows from the symmetry of $R(\tau)$, as explained above, while for small $\tau T$ the effects of excited states enter.
 As the temperature is increased, the ratio decreases nearly uniformly across the entire time range and is close to zero in the quark-gluon plasma, the signal of parity doubling.

\begin{figure}[t]
\begin{center}
\includegraphics[width=0.98\columnwidth]{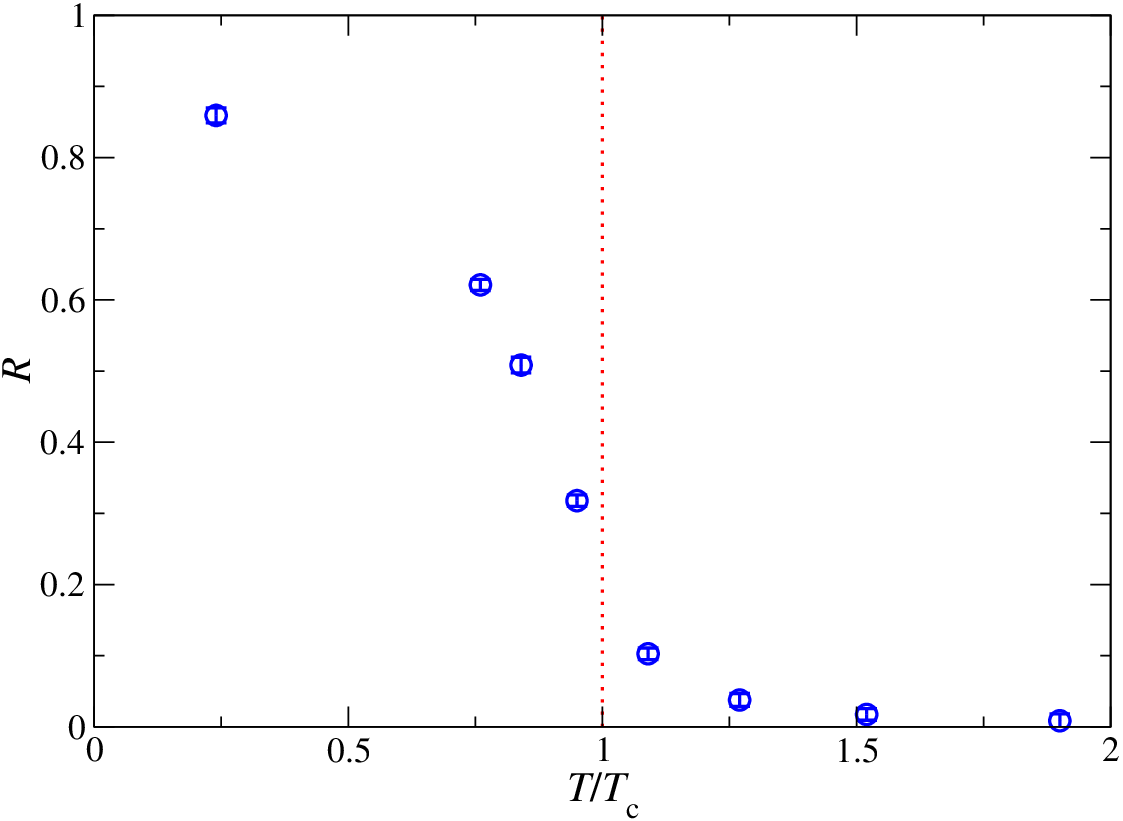}  
 \caption{Averaged ratio $R$ in Eq.\ (\ref{eq:R})  as a function of $T/T_c$.
  }
\label{fig:R}
\end{center}
\end{figure}

\begin{figure}[t]
\begin{center}
\includegraphics[width=0.98\columnwidth]{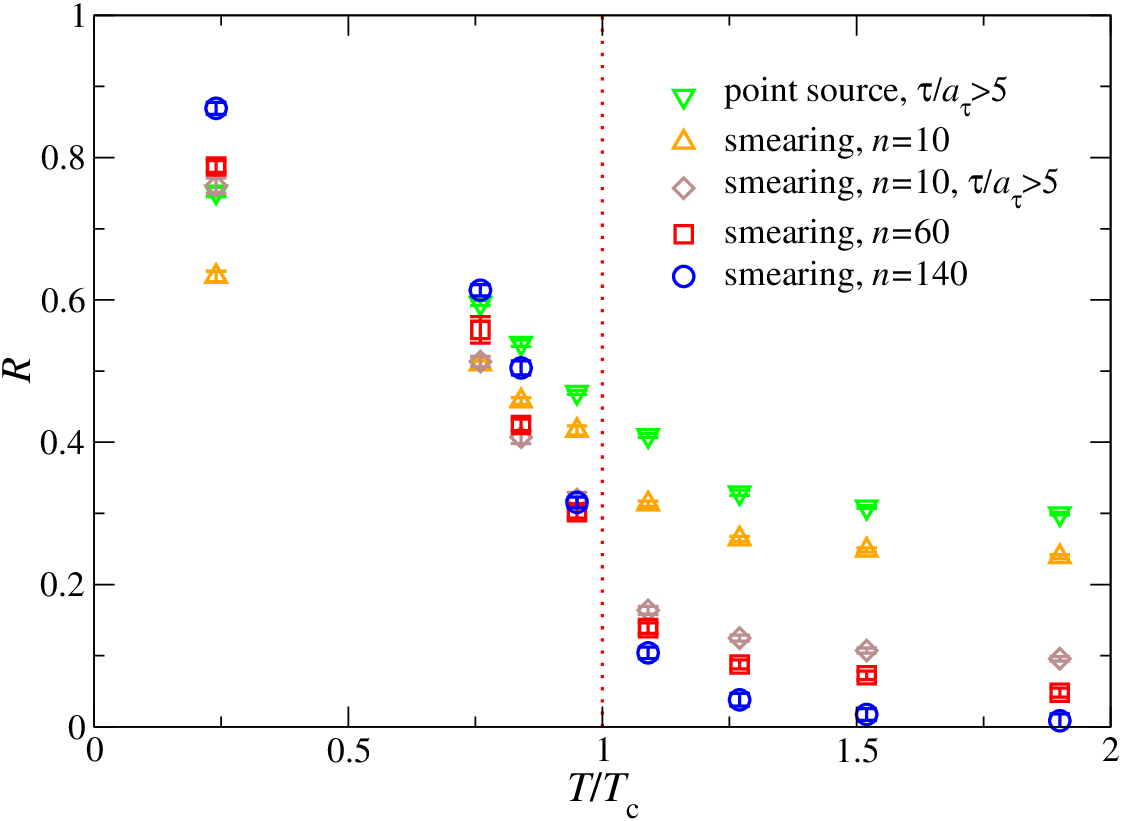}  
 \caption{As in Fig.\ \ref{fig:R}, using point sources and different levels of smearing.
   }
\label{fig:Rsmear}
\end{center}
\end{figure}

In order to quantify the parity degeneracy further, we consider the average ratio $R$, defined as
\be
\label{eq:R}
R = \frac{\sum_{n=1}^{\half N_\tau-1} R(\tau_n)/\sigma^2(\tau_n)}{\sum_{n=1}^{\half N_\tau-1} 1/\sigma^2(\tau_n)}, 
\ee
where $\sigma(\tau)$ denotes the error in $R(\tau)$ and $\tau_n=na_\tau$.
Again, the natural value of $R$ is close to but below unity in the chirally broken phase and close to zero in the parity-doubling phase.
The results for $R$ are shown in Fig.\ \ref{fig:R}. We observe a clear crossover behaviour from a nonzero value below $T_c$ to a value close to zero above $T_c$. 
The transition coincides surprisingly well with the transition to the deconfined phase, which is based on the behaviour of the Polyakov loop \cite{Aarts:2014nba}. Since deconfinement and chiral symmetry restoration are expected to occur around the same temperature  \cite{Borsanyi:2010bp,Bazavov:2011nk}, the observed parity doubling can hence be explained through the restoration of chiral symmetry in the quark-gluon plasma.

In order to achieve these results, we found that it is essential to suppress excited states at early Euclidean times, either by using smeared sources and sinks, or by considering only a restricted time interval. This is in particular pertinent for the Wilson-clover lattice fermions we use: 
since the Wilson term violates chiral symmetry and becomes relevant at larger energy scales, parity doubling of higher excited states cannot be expected. This is demonstrated in Fig.\ \ref{fig:Rsmear}, where we show the dependence of $R$ on various sources and sinks. Point sources (no smearing) couple strongly to short-distance states, for which chiral symmetry remains explicitly broken by the Wilson mass term. This can be partially taken into account by excluding the first few time slices in the analysis and considering only $\tau/a_\tau \gtrsim 5$. Applying a small number of smearing steps ($n=10$), but preserving all time slices, has approximately the same effect, while also excluding the early-time interval improves the signal considerably. Finally, applying more smearing steps ($n=60, 140$)  allows  for the signal of parity doubling to fully emerge. 
We note that since chiral symmetry breaking for Wilson fermions at large energy scales is a lattice artefact, it would be interesting to repeat this analysis using chirally symmetric fermions (see e.g.\ Ref.\ \cite{Sasaki:2001nf} for a study of parity partners using domain-wall fermions at zero temperature), especially in the chiral limit.
Finally, we have also verified that using different sources (with the same quantum numbers) yields qualitatively the same result  \cite{prep}.

\section{Summary}

 We have carried out a study of nucleons at finite temperature using lattice QCD simulations with $N_f=2+1$ dynamical flavours over a range of temperatures. We found approximate parity doubling, intimately linked with the transition to the deconfined phase. We found that the positive-parity nucleon mass is largely independent of the temperature in the hadronic phase, while the negative-parity channel shows clear temperature dependence already below $T_c$. In the context of heavy-ion collisions, we note that the temperature independence of the nucleon justifies treating the proton as unmodified by the thermal medium, while a temperature-dependent $N^*$ mass may have implications for heavy-ion phenomenology \cite{Shen:2014vra}.

As an outlook \cite{prep}, we note here that we are currently considering baryons containing strange quarks, with the aim of determining the effect from the explicit breaking by the larger strange quark mass. 
We note that the role of excited states will be of particular interest when comparing nonchiral fermions, such as in this paper, with chiral fermions, such as domain wall fermions. 
Finally, it will also be interesting to analyse our results in terms of baryon spectral functions, using Eq.\ (\ref{eq:spec}), and to address them in terms of effective models, such as the one proposed in Ref.\ \cite{Detar:1988kn} (see also Ref.\ \cite{Takahashi:2008fy} for a lattice study of the axial charge of negative-parity nucleons in this context).

\acknowledgments 

We thank STFC, the Royal Society, the Wolfson Foundation and the Leverhulme Trust for support. 
 This work used the DiRAC Blue Gene Q Shared Petaflop system at the University of Edinburgh, operated by the Edinburgh Parallel Computing Centre on behalf of the STFC DiRAC HPC Facility (www.dirac.ac.uk). This equipment was funded by BIS National E-infrastructure capital grant ST/K000411/1, STFC capital grant ST/H008845/1, and STFC DiRAC Operations grants ST/K005804/1 and ST/K005790/1. DiRAC is part of the National E-Infrastructure.
Additional computational resources were provided through HPC Wales and ICHEC.

\end{document}